\documentclass[%
reprint,
 amsmath,amssymb,
 aps,
prb,
floatfix,
showkeys
]{revtex4-2}

\usepackage{graphicx}% Include figure files
\usepackage{dcolumn}% Align table columns on decimal point
\usepackage{bm}% bold math
\usepackage{hyperref}% add hypertext capabilities
\usepackage{xfrac}
\usepackage{color}

\begin{document}

%\title{External boundary sensitivity of phononic bound states in the continuum}
%\title{Sensitivity to the boundary condition: from quasi-normal modes of open media to bound states in the continuum}
\title{Identifying bound states in the continuum by their boundary sensitivity}
\author{Vincent Laude}
\email{vincent.laude@femto-st.fr}
\author{David Röhlig}
\affiliation{%
Université Marie et Louis Pasteur, CNRS, institut FEMTO-ST, 25000 Besançon, France
}%

\date{\today}

\begin{abstract}
    We introduce a method for effectively identifying bound states in the continuum (BICs) – notably without computing the imaginary part of the eigenvalues – thereby simplifying the modeling and potentially reducing computation time.
    In real, open, physical systems, wave decay must be taken into account. This phenomenon is captured by complex-valued solutions of the harmonic wave equation, the so-called quasi-normal modes (QNMs).
    BICs, however, constitute a limiting class of solutions that do not radiate energy to infinity and are therefore, by their very nature, insensitive to the region surrounding the physical structure. Building on this observation, we identify BICs by varying the external boundary conditions that close the simulation cell in the far field; the resulting behavior is displayed in the form of spectral histograms.
    We demonstrate the effectiveness of this procedure by comparing it with conventional QNM analysis employing perfectly matched layers. Two representative examples are considered: a periodic system of permeable inclusions supporting guided Rayleigh–Bloch waves, and a whispering-gallery resonator constructed from this configuration. Finally, we provide a mathematical explanation for the method’s validity by deriving integral reciprocity statements.
\end{abstract}

%Use showkeys class option if keyword
                              %display desired
\maketitle

\section{\label{sec:level1}Introduction}
The physical description of resonators radiating in an open medium has a long history \cite{purcellPR1969} and is not only relevant to numerous physical situations in photonics \cite{sauvanPRL2013,doostPRA2014} and phononics \cite{colombiSR2016,benchabanePRA2017,landiPRL2018,schmidtPRL2018,laudePRB2018,raguinNC2019,benchabanePRA2021,elsayedPRR2020}, but applies to all quantum and classical wave problems.
Intuitively, the physical situation can be illustrated with the example of musical instruments, for instance the vibrating strings of guitars or pianos.
If idealized as being perfectly clamped at both ends, a string would, in principle, oscillate indefinitely at a single pure frequency, exhibiting no temporal decay; such a solution is typically referred to as a normal mode (NM). Yet precisely because it would radiate no energy, its tone could not reach an audience. In practice, however, the string vibration couples to the elastic waves of the body of the instrument, through the anchors fixing its ends. The energy contained in the string's motion is then radiated in the form of acoustic pressure waves in the surrounding atmosphere, thereby becoming audible to listeners. Within this framework, the temporal damping
of the vibration is directly correlated with its coupling strength to the radiation medium.
A related quantity is the quality factor~$Q$, evaluating the number of oscillations a resonator undergoes before its amplitude decays by a factor of $\exp(-\pi)$. From a physical perspective, assessing 
not only the resonance frequency but also the quality factor of each of the eigenmodes of a resonator is essential; the latter is related to the imaginary part of the eigenfrequency by $Q =\Re(f) / [2\Im(f)]$.

However, radiation loss in open systems -- where waves may escape toward infinity -- poses subtle mathematical problems and has recently led to the adaptation of the concept of a quasi-normal mode (QNM) to classical waves~\cite{chingRMP1998,sauvanPRL2013,torresPRL2020,wu2021nanoscale,laude2023quasinormal}.
Cavities and waveguides are usually characterized by their NMs defined under closed boundary conditions; in the absence of loss, the associated eigenvalue problem is Hermitian, yielding real positive eigenfrequencies, infinite quality factors, and orthogonal eigenvectors.
These properties are lost when considering QNM of open systems: eigenfrequencies are complex-valued even in the absence of material loss, and eigenmodes are not orthogonal, although their associated wave fields can still be normalized.
Closely related to the QNM concept is the notion of a bound state in the continuum (BIC) \cite{stillinger1975bound,hsu2016bound,wang2023brillouin}, which has garnered considerable attention in recent years. A BIC is a resonant state that remains completely decoupled from radiation modes, thus implying, in principle, an infinite quality factor.
BICs may arise under specific conditions, for example, for sound or water waves inside a waveguide for reasons of symmetry \cite{linton2007embedded} or in open periodic systems \cite{chaplain2025acoustic}.
In situations where the quality factor of a resonant eigenstate is extremely large yet finite, the term quasi-BIC aligns with the QNM-based mathematical description.

\begin{figure*}[t]
\includegraphics[width=0.85\textwidth]{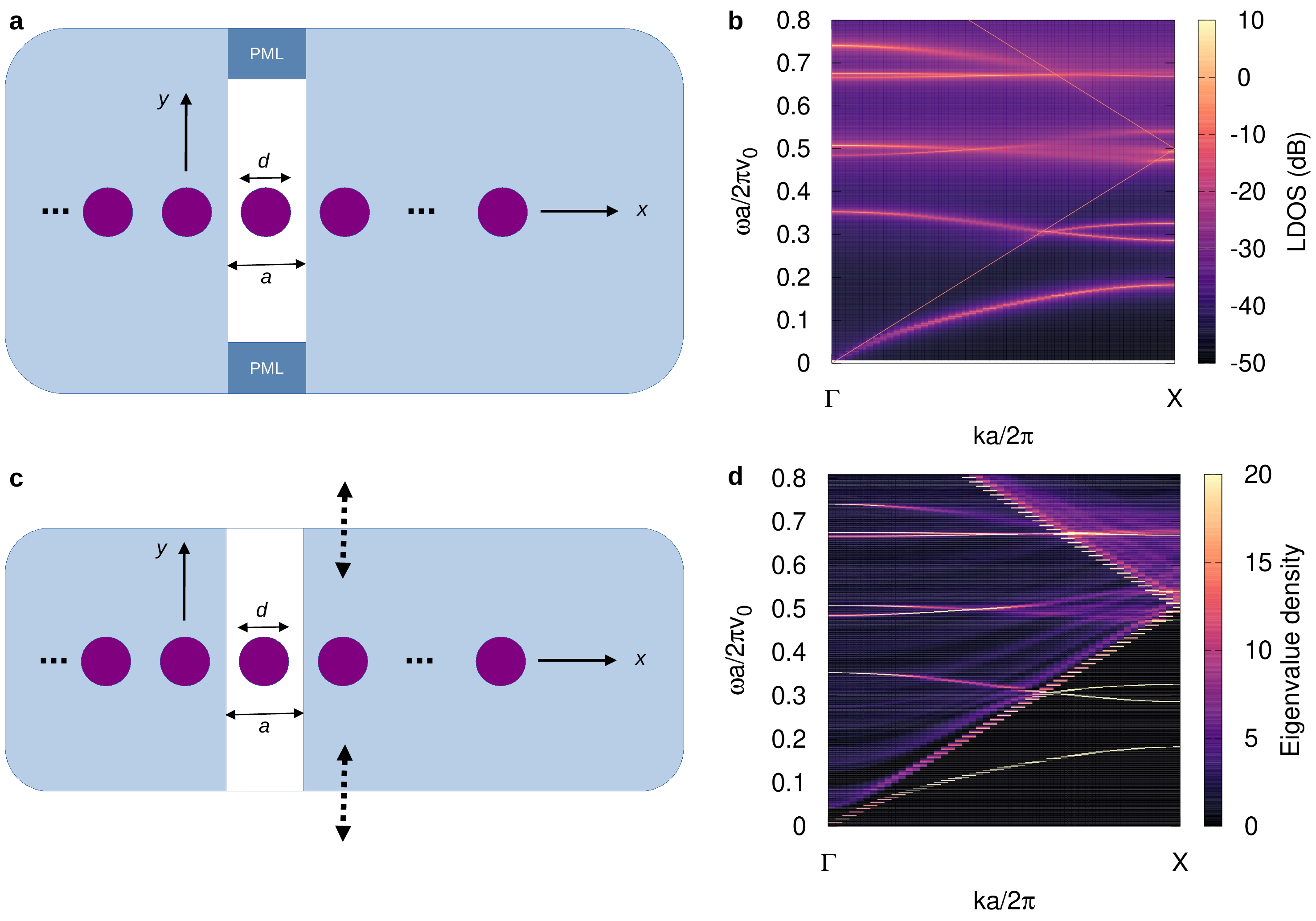}
\caption{A linear periodic chain of inclusions in a host material supporting the propagation of Rayleigh-Bloch waves \cite{laudePRB2025}. The inclusion diameter is $d$ and $a$ denotes the lattice constant. Inclusions have an elastic modulus contrast $B_2/B_1 = 1/16$ relative to the surrounding material and the same mass density.
(a) Surrounding the domain of computation with a perfectly matched layer (PML) enables the computation of quasi-normal modes.
(b) The resolvent band structure computed using the stochastic excitation method~\cite{laudePRB2018} then gives a map of the complex dispersion relation in the first Brillouin zone.
(c) Enclosing the same domain as in panel (a) using specific boundary conditions, here of the Neumann type, leads to the definition of normal modes that depend on the position of the boundary closing the waveguide.
(d) The spectral histogram is then obtained by computing classical band structures for $H=101$ cell heights ranging from $4a$ to $8a$.}
\label{fig1}
\end{figure*}

In this work, we present a new approach for effectively identifying bound states in the continuum (BICs) -- notably, without calculating the imaginary part of the eigenvalue. The idea rests on the physical property of a genuine BIC: it should be insensitive to its far-field environment. Otherwise, its eigenfrequency would change as the spatial position of the external boundary is varied, revealing a radiative leakage. Reciprocally, if the resonance is truly insensitive to the enclosing boundaries, its frequency will remain invariant and thus accumulate within a spectral histogram in the manner of a density of states. 
As we demonstrate, this observation yields a practical numerical scheme based on the physical properties of BICs. We obtain spectral histograms by iteratively varying the boundary position, which practically enables parallel processing.
We substantiate the foundations of this approach by deriving integral reciprocity statements; its effectiveness is illustrated in the setting of two related physical systems: a recently introduced linear chain of slow inclusions~\cite{laudePRB2025}, and its extension to a segmented whispering gallery resonator supporting Bloch waves~\cite{maling2016whispering}.
In the first system, the interplay of symmetry and periodicity leads to the emergence of BICs at certain points of the first Brillouin zone. For the second, recent works by Martí-Sabaté et al. have analyzed the existence of quasi-BICs and related their resonant frequency to the dispersion of Rayleigh-Bloch (RB) waves guided by the equivalent linear chain~\cite{martiPRR2023,martiCP2024}.
The fact that a linear periodic chain of inclusions supports BICs, whereas the analogous curved periodic chain supports only quasi-BICs, points to a mechanism of symmetry breaking that accounts for the emergence of radiation loss.

\section{Band histogram for Rayleigh-Bloch waves}

In this section, our aim is to reproduce the results of a previous study on RB acoustic waves~\cite{laudePRB2025} – yet now with the proposed method.
The system under consideration is a linear chain of slow inclusions, separated by a lattice constant~$a$ and possessing diameters of $0.6a$, embedded within a host medium, as depicted in Fig.~\ref{fig1}.
The elastic modulus contrast between inclusions and the surrounding material is $B_2/B_1 = 1/16$ whereas the mass densities are equal, $\rho_2/\rho_1 = 1$.
The resolvent band structure \cite{laudePRB2018} shown in Fig.~\ref{fig1}b is obtained by solving the acoustic Helmholtz equation when the unit cell is surrounded by a perfectly matched layer (PML), as depicted in  Fig.~\ref{fig1}a.
Three BICs are observed at the $\Gamma$ point and another BIC for $k a / (2\pi)=0.172$ -- see Ref.~\cite{laudePRB2025} for more details.

\begin{figure*}[th]
\centering
\includegraphics[width=0.8\textwidth]{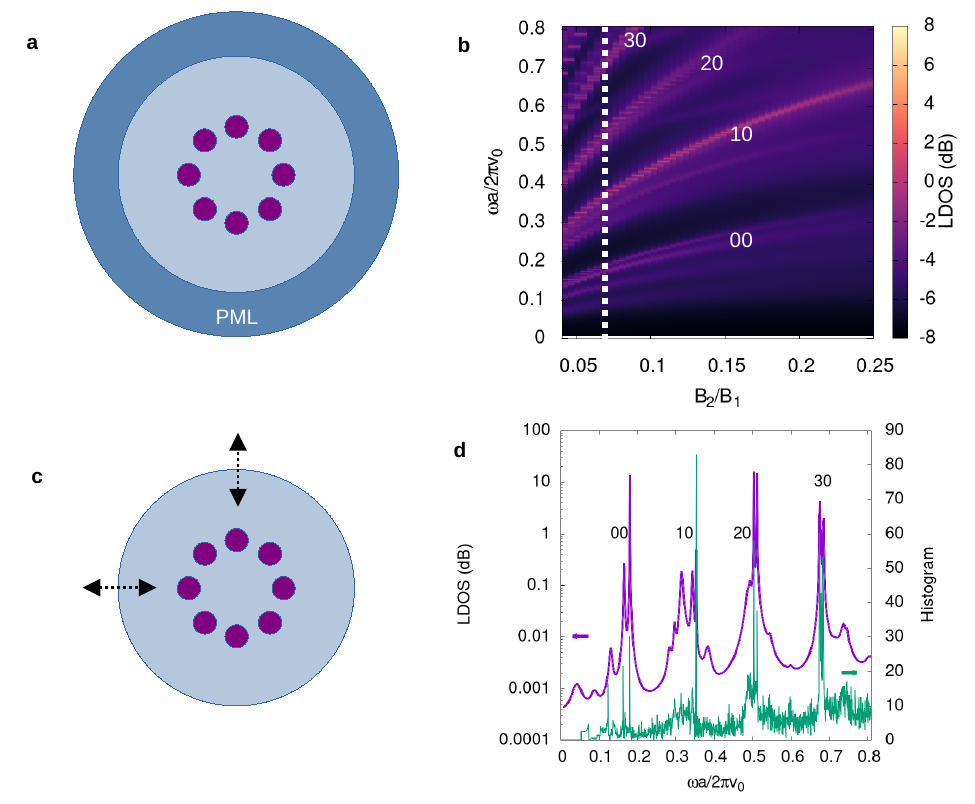}
\caption{Whispering gallery modes as quasi-normal modes (QNMs).
(a) A sequence of eight inclusions ($N=8$), each with a diameter of $0.6a$, forms a whispering gallery resonator subject to wave radiation. A perfectly matched layer (PML) emulates radiation and allows for the computation of QNMs of the resonator.
(b) The local density of states (LDOS) computed using the stochastic excitation method, as a function of frequency and the elastic contrast between the inclusions and the host medium shows groups of QNMs that can be labeled with two indices $mn$. The azimuthal index $m$ varies from 0 to 3, and the radial index $n=0$ is fixed.
The vertical dotted line is placed at $B_2/B_1 = 1/16$.
(c) Enclosing the same domain as in panel (a) using a Neumann boundary condition leads to the definition of normal modes that depend on the position of the boundary closing the computation domain.
(d)
The local density of states of quasi-normal modes is shown with a magenta line for contrast $B_2/B_1 = 1/16$ and is compared with the spectral histogram of normal modes, shown with a green line. The latter was obtained by the sum of normal distributions of Eq. (\ref{eq:gaussian}) for $H=101$ cell diameters ranging from $2L$ to $4L$.
}
\label{fig2}
\end{figure*}

Instead of considering the resolvent band structure, we next compute the classical band structure where the cell is enclosed by Neumann boundaries of varying width.
We form the frequency histogram as a function of $k$ and look at the accumulation points in the dispersion space.
More precisely, we solve repeatedly for the classical band structure, $\omega(k; \partial \Omega)$, varying the position of the external boundary $\partial \Omega$. This is done within the far field, where the evanescent components of the solution no longer contribute meaningfully. Were the cell size chosen within the near-field domain, this would manifest itself in a blurred spectral histogram. In this sense, the histogram of a BIC further provides a quantitative diagnostic tool for distinguishing near from far field.
Considering~$H$ different positions of the external boundary and~$P$ eigenfrequencies per wavevector, the histogram in dispersion space $(k, \omega)$ is constructed by accumulating eigenfrequencies in intervals of width $\delta\omega$, or
\begin{equation}
h(\omega) = \sum_{i=1}^{H} \sum_{p=1}^{P} \Pi \left( \frac{\omega - \omega_p^i}{\delta \! \omega} \right)
\end{equation}
where $\Pi(x)$ is the rectangle function equal to $1$ if $-1/2 \leq x \leq 1/2$ and $0$ otherwise. Intuition indeed suggests that the normal frequency of a mode that depends on the external boundary conditions changes as the latter are varied. Hence, if the sensitivity is high, frequencies are expected to disperse broadly over many intervals. Conversely, if the sensitivity to the boundary condition is low, the frequencies tend to accumulate within a narrow interval.

Using the lattice constant~$a$ as the primary scale, we change the waveguide width from a lower bound of $4a$ – which is well above the wavelengths of interest after the first band folding – up to $8a$. $H=101$ classical band structures are calculated, each with $P=32$ eigenfrequencies (which is more than enough to cover the displayed spectral range); $\delta \omega$ is set such that~$12000$ frequency points are used. The resultant spectral histogram, shown in Fig.~\ref{fig1}d, admits a direct comparison with the resolvent band structure; bands manifest themselves at the same locations in dispersion space.
As a notable difference between the two simulation techniques, the spectral histogram is computed for NMs, whereas the resolvent band structure is computed for QNMs.
Some of the NMs are seen to concentrate along the sound line, which is consistent with bulk waves being independent of external boundaries.
More interestingly, NMs also concentrate along the quasi-normal bands that were previously identified~\cite{laudePRB2025}.
The highest concentrations are observed for waves guided below the sound line, which are by definition evanescent normally to the waveguide axis and hence rather insensitive to the position of the external boundaries.
Significantly, NMs also concentrate at the BIC points that lie within the sound cone, both at the high symmetry points of the first Brillouin zone and within it.
In partial conclusion, although spectral histograms do not allow the estimation of radiation loss, they nonetheless provide meaningful insight into the presence of BICs and quasi-BICs. Even without direct access to the imaginary part, reliable conclusions can be drawn from the sensitivity of normal modes to the external boundary condition.

\section{Whispering gallery modes as quasi-normal modes}

We next consider a circular periodic sequence of $N$ disk inclusions with a diameter of $0.6a$, as depicted in Fig.~\ref{fig2}a in the case $N=8$.
The circle formed by the inclusions has a radius $L$, measured at the centers of the disks. The angular repetition of the disks is therefore $2\pi/N$. 
The disk centers have coordinates
\begin{eqnarray}
    x_i = L \cos\left(\phi_i \right), \quad
    y_i = L \sin\left(\phi_i \right), \\
    \phi_i = \dfrac{2\pi}{N} i, \quad
    \forall i=0, N-1. \nonumber
\end{eqnarray}
The radius $L$ of the sequence is thus related to the lattice constant $a$ of the Rayleigh-Bloch chain as
\begin{equation}
a =  \frac{2\pi}{N}L.
\end{equation}
By fixing the radius of the inclusions and the distance between them, resonances of the whispering gallery appear at about the same frequency as the corresponding Rayleigh-Bloch waves of the linear chain.
However, because of the curvature of the chain and its finiteness, some degree of leakage is expected to occur, causing BICs to become quasi-BICs.

Since the structure considered is a resonator in an open, unbounded host medium, we expect it to show a sequence of QNMs whose imaginary part of the frequency measures radiation loss.
The resonance frequencies -- connected to the real part of the complex frequency of each QNM -- can be explored using the stochastic excitation of the inclusions.
The result is displayed in Fig.~\ref{fig2}b as a function of the elastic modulus of the inclusion, normalized to that of the host medium, again for equal mass densities.
The plot reveals distinct groups of QNMs whose resonance frequencies decrease continuously as the modulus ratio is reduced.
As we illustrate below, those modes are hybridizations of the QNMs of a single
inclusion.
As discussed earlier~\cite{laudePRB2025}, the shape of a QNM for a disk can be systematically ordered and classified by means of the azimuthal index~$m$ and the radial index~$n$. The same scheme can be applied to the entire disk sequence and is used to label the groups of QNMs appearing in Fig.~\ref{fig2}b.

\begin{figure}[t]
\centering
\includegraphics[width=\columnwidth]{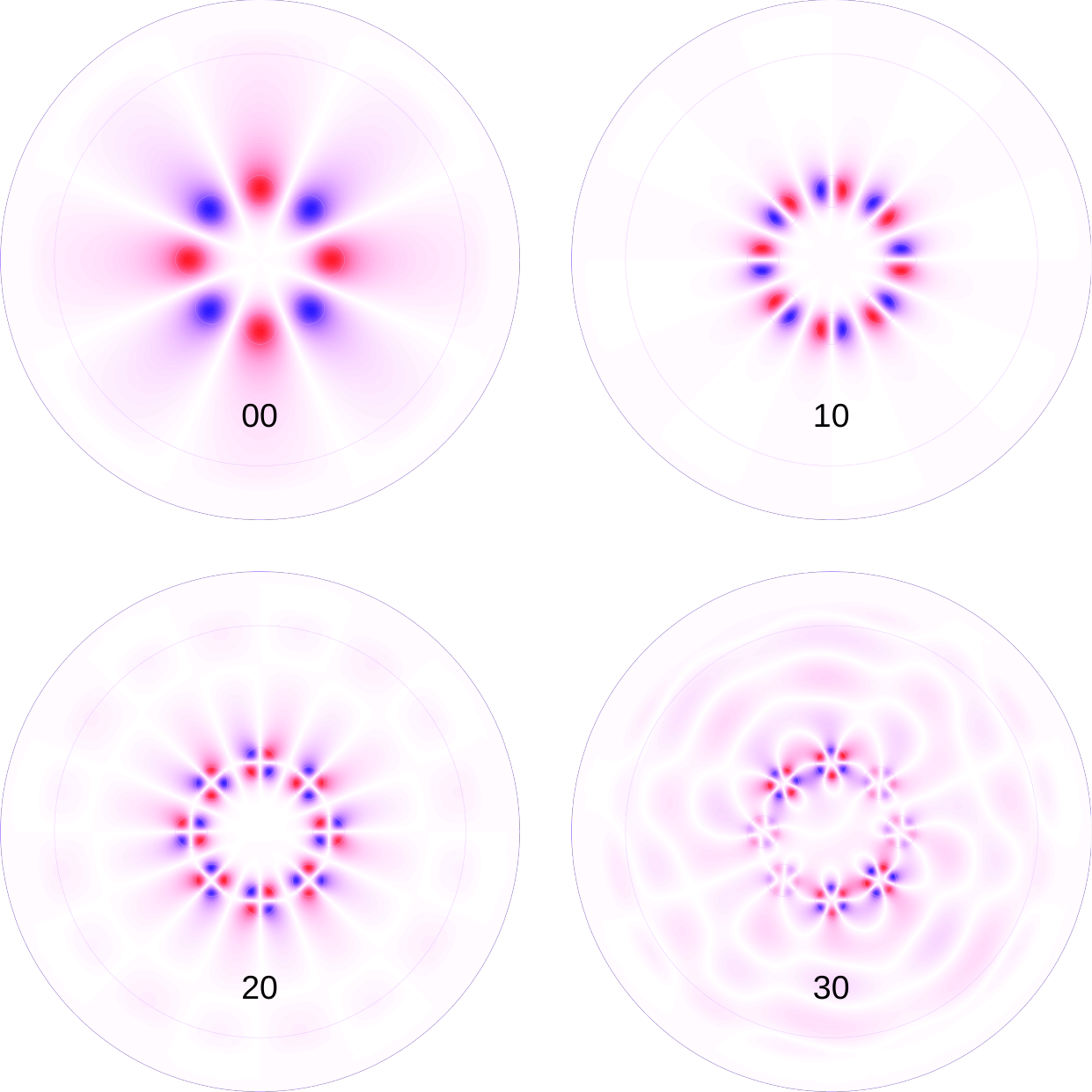}
\caption{Quasi-normal, whispering-gallery, modes of a circular sequence of 8 slow inclusions with diameters of $0.6a$ for a modulus ratio of $B_2/B_1 = 1/16$. The symmetry of each mode is classified by $mn$, where $m$ denotes the azimuthal index and $n$ the radial number.}
\label{fig3}
\end{figure}

\begin{table}[h]
\caption{Properties of the quasi-normal, whispering-gallery, modes in Fig.~\ref{fig3}.}
\centering
\begin{tabular}{rcr}
$mn$ & \quad $\omega a / 2 \pi v_0$ & $Q$ \\ \hline
00 & 0.181 & 472 \\
10 & 0.353 & 160000 \\
20 & 0.512 & 1535 \\
30 & 0.676 & 1354
\end{tabular}
\label{tab1}
\end{table}

\begin{figure}[t]
\centering
\includegraphics[width=\columnwidth]{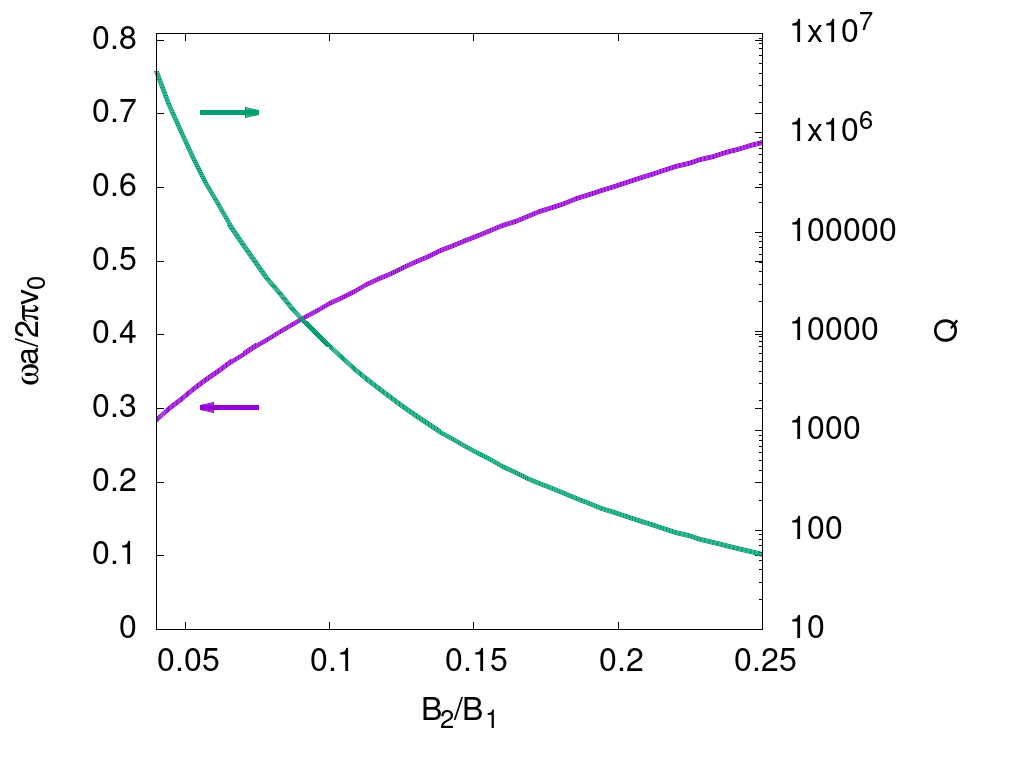}
\caption{
Evolution of both the frequency and the quality factor~$Q$ of quasi-normal mode~$mn=10$ as a function of the elastic modulus ratio $B_2/B_1$.}
\label{fig4}
\end{figure}

The definition of a spectral histogram by a simple accumulation inside sequential intervals introduces some degree of arbitrariness, as the frequency of a given NM jumps discretely from one interval to the next according to the chosen frequency resolution.
We can instead smooth the histogram by affecting a normal, or Gaussian, probability distribution to each frequency sample.
Specifically, we define
\begin{equation}
g(f) = \frac{1}{\delta \! \omega}\sum_{i=1}^{H} \sum_{p=1}^{P} \exp\left[ - \frac{(\omega - \omega_p^i)^2}{2 \, \delta \! \omega^2} \right].
\label{eq:gaussian}
\end{equation}
The resulting spectral histogram for $B_2/B_1 = 1/16$ is presented in Fig.~\ref{fig2}d. The Neumann boundary was displaced sufficiently far from the structure – specifically, across $H=101$~cell diameters ranging from~$2L$ to~$4L$; $\delta \! \omega$ was set such that it equals $1/6000$ of the displayed spectral range.
With this frequency resolution, sharp peaks clearly appear for the modes with the lowest radiation loss, or the highest quality factor.
The spectral histogram can be directly compared with the density of states  also shown in Fig.~\ref{fig2}d.
Maxima as a function of frequency coincide better for the sharpest peaks in the density of states, hence for the largest quality factors.
This is again a manifestation of the relation of the sensitity of NMs to the boundary condition with radiation loss; a mathematical derivation of this relation is given in the next section.
Before moving on to this discussion, we display in Fig.~\ref{fig3} the modal shapes for QNMs with the highest quality factor in each group of eigenmodes.
Owing to their large~$Q$ values listed in Table~\ref{tab1}, these modes may be interpreted as quasi-BICs. They were calculated using an algorithm to extract the QNM closest to a given target frequency, as described in Ref.~\cite{laude2023quasinormal}.
The azimuthal symmetry of the original modes of the disk, labeled by azimuthal index $m$, is clearly preserved in the modes of the circular chain.
For completeness, the variation with the elastic modulus ratio of the best QNM, namely $mn=01$, is shown in Fig.~\ref{fig4}; the quality factor~$Q$ varies  acutely with the inverse of the elastic modulus contrast and can exceed $10^6$.

\section{Reciprocity relations}

Can we relate QNM and NM by establishing a mapping between them – thereby providing conceptual support for the numerical results presented before?
In general, the weak form of the source-free scalar Helmholtz equation takes the following structure: 
\begin{eqnarray}
\int\limits_\Omega \nabla v \cdot \left( \alpha \nabla u \right)
- \omega^2 \int\limits_\Omega v \beta u & = & \oint\limits_{\partial \Omega} v \left( \alpha \nabla u \right) \cdot \mathbf{n}. \qquad
\end{eqnarray}
It remains valid for an arbitrary domain of integration~$\Omega$; thus the external boundary $\partial \Omega$ with its associated normal vector $\mathbf{n}$ can be placed arbitrarily as well. The quantities~$\alpha$ and~$\beta$ are connected to the relevant material parameters – for instance, permeability and permittivity in electromagnetics, or bulk modulus and mass density in the context of acoustics (we implicitly used $\alpha=B^{-1}$ and  $\beta=\rho^{-1}$ in the previous sections).

Let $u_1$ be a QNM with eigenfrequency $\omega_1^2 \in \mathbb{C}$. It is obtained by enclosing the finite domain $\Omega_0$ that contains the resonator with a surrounding PML labeled $\Omega_\text{P}$. The integration domain is therefore given by $\Omega = \Omega_0 \,\cup\, \Omega_\text{P}$. Since the PML is itself terminated by a Dirichlet boundary condition, the corresponding boundary integral vanishes. Then, the QNM satisfies
\begin{eqnarray}
\int\limits_{\Omega_0} \nabla v \cdot \left( \alpha \nabla u_1 \right) - \omega_1^2 \int\limits_{\Omega_0} v \beta u_1 \; & = &  \\
- \int\limits_{\Omega_\text{P}} \nabla v \cdot \left( \alpha \nabla u_1 \right) + \omega_1^2 \int\limits_{\Omega_\text{P}} v \beta u_1. &&\nonumber
\end{eqnarray}
When inserting the complex conjugate of the field by setting $v=u_1^*$, this equation implies that the difference of kinetic and elastic energies -- the Lagrangian $L$ -- of the QNM inside $\Omega_0$ is balanced by the same quantity evaluated inside $\Omega_\text{P}$.
Alternatively, by changing the integration domain to only $\Omega = \Omega_0$, we can write
\begin{eqnarray}
\;\;\,\int\limits_{\Omega_0} \nabla v \cdot \left( \alpha \nabla u_1 \right) - \omega_1^2 \int\limits_{\Omega_0} v \beta u_1\; & = & \oint\limits_{\partial \Omega} v \left( \alpha \nabla u_1 \right) \cdot \mathbf{n}. \qquad
\label{eq:qnm}
\end{eqnarray}
Incidentally, the last identity proves that
\begin{eqnarray}
- \int\limits_{\Omega_\text{P}} \nabla v \cdot \left( \alpha \nabla u_1 \right) + \omega_1^2 \int\limits_{\Omega_\text{P}} v \beta u_1 & = & 
\oint\limits_{\partial \Omega} v \left( \alpha \nabla u_1 \right) \cdot \mathbf{n}. \qquad
\end{eqnarray}
When setting $v=u_1^*$ again, the expression can be interpreted as follows: the power flux radiated by the QNM is balanced by the Lagrangian evaluated within the PML.
In contrast, let us now consider a NM~$u_2$ with eigenfrequency~$\omega_2^2 \in \mathbb{R}$; it is simply given by
\begin{eqnarray}
\int\limits_{\Omega_0} \nabla v \cdot \left( \alpha \nabla u_2 \right) - \omega_2^2 \int\limits_{\Omega_0} v \beta u_2 & = & 0,
\label{eq:neumann}
\end{eqnarray}
%with $\omega^2 \in \mathbb{R}$, 
since the boundary integral vanishes because of the Neumann boundary condition applied on $\partial\Omega$.
Notably, the QNM $u_1$ and the NM $u_2$ differ whenever there is radiation through the external boundary.

Next, we combine the QNM Eq.~(\ref{eq:qnm}) using the test function ~$v = u_2$, and, reciprocally, the NM Eq.~(\ref{eq:neumann}) with the test function $v = u_1$.
Subtracting them and using the symmetry of the stiffness operator we obtain
\begin{eqnarray}
(\omega_2^2 - \omega_1^2) \int\limits_{\Omega_0} u_2 \beta u_1 \; & = & \oint\limits_{\partial\Omega} u_2 \left( \alpha \nabla u_1 \right) \cdot \mathbf{n}.
\label{eq:main}
\end{eqnarray}
This relation constitutes our main theoretical result and may be regarded as a form of reciprocity statement.
If the QNM is known, it reveals how the eigenvalue $\omega_2^2$ of the NM will depart from the eigenvalue $\omega_1^2$ of the QNM: the amount of deviation is proportional to the QNM’s radiated power flux projected onto the eigenvector of the NM. This relation is somehow re-entrant, so we may use the following zero-order approximation instead.
Under the perturbation caused by imposing a Neumann, rather than the radiation boundary condition (or its PML approximation), the eigenvector is expected to vary less quickly than the corresponding eigenvalue.
Hence, using the zero-order approximation $u_2 \approx u_1$, we get an order-one perturbation estimate for $\omega_2^2$ as
\begin{eqnarray}
(\omega_2^2 - \omega_1^2) \int\limits_{\Omega_0} u_1 \beta u_1 \; & \approx & \oint\limits_{\partial\Omega} u_1 \left( \alpha\nabla u_1 \right) \cdot \mathbf{n}.
\label{eq:main2}
\end{eqnarray}
Of course, if the QNM has decreased sufficiently at the external boundary $\partial\Omega$, then the boundary integral tends toward zero and $\omega_2^2 \approx \omega_1^2$.
In this case, the NM's eigenfrequency corresponds to a BIC – a QNM of infinite quality factor – and thus coincides with the QNM frequency, which is consequently real-valued. As a result, a straightforward criterion for determining whether a QNM constitutes a BIC is
\begin{eqnarray}
\oint\limits_{\partial\Omega} u_1 \left( \alpha \nabla u_1 \right) \cdot \mathbf{n} = 0,
\end{eqnarray}
that is, the requirement that the QNM satisfy the Neumann boundary condition in an integral sense.

Finally, let us relate the imaginary part of the complex QNM eigenfrequency with the normal mode shape.
We assume that the quality factor is large enough that $\omega_1 \approx \omega_2 (1 - \imath \gamma)$, i.e. that the real part of $\omega_1$ is close to $\omega_2$. Then $\omega_2^2 - \omega_1^2 \approx 2 \imath \gamma \omega_2^2$.
When combining Eq.~\eqref{eq:main2} with Eq.~\eqref{eq:neumann} for $v=u_2^*$ we obtain
\begin{eqnarray}
\gamma = \frac{1}{2} \Im\left[ \frac{\oint\limits_{\partial\Omega} u_1 \left( \alpha\nabla u_1 \right) \cdot \mathbf{n}}{\int\limits_{\Omega_0} u_1 \beta u_1} \right] \left[ \frac{\int\limits_{\Omega_0} u_2^* \beta u_2}{\int\limits_{\Omega_0} \nabla u_2^* \cdot \left( \alpha \nabla u_2 \right)} \right] . \quad
\end{eqnarray}
The choice of the test function $v=u_2^*$ is made so that the integrals defining $\omega_2^2$ are real and positive but remains otherwise arbitrary.

\section{Discussion}
In summary, we have presented a conceptually simple yet effective approach for identifying BICs without recourse to calculate the imaginary parts of the eigenvalues. By viewing BICs as inherently insensitive to the exterior of the physical structure, we exploit variations in the external boundary conditions to reveal their presence through regions of frequency accumulation in spectral histograms. Due to the presumed far-field decoupling, high-wavelength solutions (e.g. in the homogenisation limit) are excluded from detection.
A mathematical justification – grounded in the pertinent integral reciprocity relations – elucidates why the method succeeds.
A comparison with conventional QNM analysis employing PMLs shows good agreement for both representative examples considered: a periodic array of slow inclusions supporting guided Rayleigh–Bloch waves, and a related whispering-gallery resonator. 
% – demonstrate the method’s applicability across different settings.
%Looking ahead, it would be of particular interest to relate the imaginary parts of the solutions to the degree of dispersion across intervals in the histograms. Further work is needed to quantitatively connect histograms to quality factors, especially if they are not too large. We leave this question to subsequent investigations.

From a computational perspective, we note that the inherently iterative scheme enables parallelization. In contrast, complex eigenvalue solvers typically exhibit strong inter-iteration dependencies, which substantially constrain efficient parallel execution. When distinct real-valued simulations are assigned to independent cores, the expected speed-up is close to linear for the parameter ranges considered in our examples.
Moreover, the efficiency of BIC detection can be tuned by adjusting the spatial variation of the boundary; a smaller number of judiciously selected computations may suffice. A geometrically scaled progression of the cell-size parameter might be preferable to a linear sequence with equidistant spacing.
In order to determine an appropriate number of simulations, one may introduce a suitable indicator together with a corresponding termination condition. In the present study, a linear progression was employed until the histogram revealed a discernible accumulation pattern. While sufficient for our purposes, it yet leaves the systematic determination of an optimal sampling strategy open to further investigation.
Finally, although the concept has been demonstrated here for the scalar wave equation, no fundamental conceptual distinction arises between scalar and vectorial wave problems. The proposed method should therefore extend naturally to vector waves. 
Taken together, these considerations establish a practical and conceptually coherent framework for the systematic detection of BICs in wave systems.

\acknowledgments
This work is dedicated to the memory of Dr. Sarah Benchabane.
It has been supported by the EIPHI Graduate school (contract "ANR-17-EURE-0002") and by the Bourgogne-Franche-Comté Region.
Finite element computations were performed with FreeFem$++$ \cite{freefem}.

\section*{Data availability statement}

Numerical data and codes to reproduce them will be made openly available.
%are openly available at \cite{RBwaves}.

\bibliography{qnm,RBwaves}% Produces the bibliography via BibTeX.

\end{document}